\UseRawInputEncoding
\documentclass[preprint]{aastex631}
\accepted{September 16, 2021}
\submitjournal{ApJ}

\shorttitle{
Footpoint signatures of  a soft X-ray loop-like microflare
}
\shortauthors{Shimizu et al.}
\begin{document}

\title{
%Time variability in chromospheric temperature of a solar plage region measured 
%with ALMA at 3 mm
%Nanoflare-class energy inputs in solar chromosphere revealed with ALMA at 3 mm
Simultaneous ALMA−Hinode−IRIS observations on 
footpoint signatures of a soft X-ray loop-like microflare %revealed with ALMA at 3mm
}

\correspondingauthor{Toshifumi Shimizu}
\email{shimizu@solar.isas.jaxa.jp}

\author[0000-0003-4764-6856]{Toshifumi Shimizu}
\affil{ Institute of Space and Astronautical Science, Japan Aerospace Exploration Agency, 3-1-1 Yoshinodai, Chuo, Sagamihara, Kanagawa 252-5210, Japan}
\affiliation{Department of Earth and Planetary Science, The University of Tokyo, 7-3-1 Hongo, Bunkyo-ku, Tokyo 113-0033, Japan}

\author[0000-0002-2350-3749]{Masumi Shimojo}
\affil{National Astronomical Observatory of Japan, 2-21-1, Osawa, Mitaka, Tokyo, 181-8588, Japan}
\affil{Department of Astronomical Science, SOKENDAI (The Graduate University for Advanced Studies), 2-21-1, Osawa, Mitaka, Tokyo 181-8588, Japan}

\author{Masashi Abe}
\affil{Department of Earth and Planetary Science, The University of Tokyo, 7-3-1 Hongo, Bunkyo-ku, Tokyo 113-0033, Japan}
\affil{ Institute of Space and Astronautical Science, Japan Aerospace Exploration Agency, 3-1-1 Yoshinodai, Chuo, Sagamihara, Kanagawa 252-5210, Japan}

%\begin{comment}
%\altaffiliation{Creator of AASTeX v6.2}
%\nocollaboration



\begin{abstract}
%Small-class 
%transient reconnection events, that is, microflares and nanoflares 
Microflares 
have been considered to be among the major energy input sources 
to form active solar corona. 
%Recent high resolution observations allow us to investigate 
%the structure of elementary energy releases involved in a single event. 
To investigate the response of the low atmosphere to events, 
we conducted an 
%Atacama Large Millimeter/sub-millimeter Array (ALMA) observation 
ALMA observation
at 3 mm coordinated with 
%Interface Region Imaging Spectrograph (IRIS) and 
IRIS and
Hinode observations, on March 19, 2017.
During the observations, 
a soft X-ray loop-type microflare (active-region transient brightening) was 
captured using Hinode X-ray telescope in high temporal cadence.  
A brightening loop footpoint is located within narrow
field of views ALMA, IRIS slit-jaw imager, and Hinode spectro-polarimeter.
Counterparts of the microflare at the footpoint were detected in Si IV and ALMA images, 
while the counterparts were
less apparent in C II and Mg II k images.  
Their impulsive time profiles exhibit the Neupert
 effect pertaining to soft X-ray intensity evolution. 
 The magnitude of thermal energy measured using ALMA
 was approximately 100 times
 % is about one hundred
smaller than that measured in the corona. 
These results suggest that impulsive counterparts can be detected
in the transition region and upper chromosphere where the plasma is thermally
heated via impinging non-thermal particles. Our energy evaluation
indicates a deficit of accelerated particles that impinge the footpoints for
a small class of soft X-ray microflares. 
 The footpoint counterparts consist of several brightening kernels, 
 %(size = 1\arcsec\ or less),
 %arc-s or less in size), 
 all of which are located in weak (void) 
 magnetic
 areas formed in patchy distribution of strong magnetic flux 
at the photospheric level.
The kernels provide a conceptual image in which
 the transient energy release occurs
at multiple locations on the sheaths of magnetic flux bundles
in the corona. 
\end{abstract}

%\keywords{Sun: chromosphere --- Sun: radio radiation --- Sun: transition region ---
%Sun: corona --- Sun: activity --- Sun: flares --- Sun: magnetic fields
\keywords{
{\it Unified Astronomy Thesaurus concepts:}
Solar physics[1476], Solar electromagnetic emission[1490], Solar activity[1475], Solar flares[1496]
}


\section{Introduction}
\label{sec:intro}

The solar corona 
%is the outer atmosphere of the Sun, which has 
is formed at more than $10^{6}$ K 
above the surface (photosphere) with a temperature of 6,000 K. 
%This temperature structure cannot be explained by the 1st law of thermodynamics. 
%The ”photosphere-corona” structure is
%generally common in late type stars, and this ”coronal heating” problem has been tackled by a lot
%of scientists with theoretical and observational approaches for a long time.
%In the solar corona, the energy flux required to maintain the $10^{6}$ K plasma 
%is estimated to be $5\times10^5$ erg cm$^{-2}$ s$^{-1}$ in the quiet area
%, i.e., the most portion of the solar surface, 
%and it is increased to $10^7$ erg cm$^{-2}$ s$^{-1}$ 
%at bright regions in X-rays, i.e., active regions. 
The corona is highly structured and has active regions with high energy input that
are observed to be bright and active in soft X-rays and high-temperature extreme ultraviolet (EUV) lines.
%In general, two groups of theoretical ideas
%have been proposed and examined: wave heating and reconnection (microflare-nanoflare) heating.
%High spatial resolution and temporal cadence observations from recent advanced instruments, such
%as telescopes onboard the Hinode spacecraft, have revealed MHD waves excited in the corona [2, 3]
%and have given great improvements on our understanding of waves as the major contributor to the
%heating of the corona in the quiet area. On the other hand, the heating in active regions still remains
%a great mystery. The active region requires the heating rate roughly 20 times higher than the quiet
%area does. Recent observations suggest that the propagating waves observed in the low atmosphere
%with the Hinode and IRIS satellites give insufficient heating rate to the corona [4]. Rather, 
%Out of a lot theoretical ideas for the energy supply, nanoflares, which is a tiny class of magnetic
%reconnection events in order of $10^{24}$ erg, could be important for maintaining active-region corona, 
%because extremely high temperature plasma ($> 8$ MK) exists in active regions [5, 6].
Small-class transient reconnection events, that is, microflares (in the order of $10^{27}$ erg) 
and nanoflares (in the order of $10^{24}$ erg),  
have been considered to be among the major energy input sources 
to form the active corona. This is because extremely high temperature plasma ($> 8$ MK) 
is observed in active regions \citep{yos96, ter12}. 
Microflares have been predominantly observed as transient brightenings in active regions
\citep{shi92} that morphologically show various brightening coronal structures 
in soft X-rays \citep{shi94}. 
Simultaneous multiple loop brightenings are more often seen than single loop brightenings
and the loops tend to brighten either from their footpoints and/or from
the the apparent contact point , followed by the brightening of the entire loops.
In such cases, the energy release site is located in the corona. 
Certain portion of the energy released in the corona is transported 
via accelerated particles and thermal conduction along the magnetic field
to the footpoints, where the chromospheric and transition region cool plasma
is transiently heated. We are yet to understand the partition of the energy;
the amount of energy transferred to the footpoints and significant heating of
the chromosphere. 
Studies have been conducted on the energy partition in flares 
(references in Section \ref{sec:partition}). 
Observational understanding of the energy partition in microflares and 
nanoflares can significantly expand the range of energy to disucss
the energy partition. This helps to understand the energy partition changes as 
a function of event energy size. 

%For major flares, we have empirical formulae to convert the soft X-ray flux to
%the total radiant energy release: $L_{total}/L_{SXR} = 15$ (Wu et al. 1986).
%Dominant is the radiance of spectral lines originating from the chromospheric
%plasma, which are transiently brightened by the energy transport from the corona
%to the chromosphere. We do not know whether this empirical ratio is valid for
%smaller energy release events, but it was used as one of the methods to estimate
%the released energy of soft X-ray microflares (Shimizu 1995). 
%however, magnetic reconnection at the chromospheric bases of coronal magnetic structures 
%is one of candidates to drive spicules \citep{tsi12}. High-speed upflows and turbulence
%are observed at the footpoints of coronal loop structures \citep{har08, mci09}. 

The energy released by small-scale transient events has been evaluated as 
thermal energy via temperature and density diagnostics using soft 
X-ray or EUV intensities. This provides frequency distribution in
microflare and nanoflare energy range  \citep[and references herein]{shi95, ash00}. 
Hard X-ray observations are also used to evaluate
non-thermal energy; however, they are solely conducted
for flares above $5 \times 10^{28}$ erg \citep{cro93}.
The hard X-ray emission originated predominantly in the accelerated particles
that impinge the lower atmosphere at the footpoints of the magnetic structure of
coronal flaring.
The Interface Region Imaging Spectrograph (IRIS)\citep{dep14}, launched in 2013,
has a high-resolution used to observe the movement of heat and energy 
in the chromosphere and lower portion of the coronal transition region.  
IRIS has a spectrograph and slit-jaw imagers to 
monitor the intensity of some wavelength bands that contain spectral lines originating
from the chromosphere and lower portion of transition region. 
Later in the decade following 2010, a new diagnostic capability for the chromosphere was incorporated
when
the Atacama Large millimeter and sub-millimeter Array (ALMA) \citep{alm16} began
solar observations. 
ALMA can perform interferometric observations of the Sun at millimeter (mm) wavelengths. 
Most of the mm emission comes from the chromosphere \citep{ver81}. 
The mm-wavelength emission is
a free-free emission (thermal bremsstrahlung) produced by free electrons scattering 
off ions without being captured. The free electrons can be assumed to be in local 
thermodynamic equilibrium: thus, the source function is equivalent to the Plank function. 
In the mm wavelengths, the Rayleigh--Jeans law is an approximation to the Plank function
that demonstrates a linear relationship between the emission intensity and 
the brightness temperature \citep{kra86}.  
This indicates that the radiation at mm wavelengths can be used as a thermometer 
to probe the chromospheric temperature. 
The thermometer provides acts as a useful tool to detect changes in the chromospheric temperature
to estimate the amount of energy transferred to the chromosphere.

In this study, we present the first example of small-scale transient brightenings produced
in the solar corona (hereafter, referred to as soft X-ray microflare) 
and their responses in the chromosphere and lower portion of 
coronal transition region.
The soft X-ray microflare was captured in high temporal cadence using Hinode 
X-ray telescope to observe the evolution of brightening loop structure in the corona
and locations of the brightening loop footpoints on the solar surface.
One footpoint was within the narrow fields of view of the 
ALMA, IRIS, and Hinode spectro-polarimeter. 
The  spectro-polarimeter provides high precision and high spatial resolution maps of
magnetic flux distributed on the photosphere, which
are helpful to understand the magnetic field configurations of the soft X-ray microflare.

%The Atacama Large millimeter and sub-millimeter Array (ALMA) \citep{alm16} is capable of
%performing interferometric observations of the Sun at mm wavelengths, allowing
%both high spatial and temporal resolution imaging observations
%of a few arcsec or higher spatial resolution \citep{shm17}.  
%The objective of this article is to present the temporal behaviors in 
%mm-wavelength data recorded with an ALMA observation in the Cycle 4 period,
%which was the first proposal opportunity for the solar community. 
This article is organized as follows. 
Section \ref{sec:obs} describes the observations and data analysis.
Section \ref{sec:results} presents the results of the data analysis.
Section \ref{sec:discuss} and Section \ref{sec:summary} present the discussion
of the results and summary, respectively. 

\section{Observations and  Data analysis} 
\label{sec:obs}

\subsection{Observations}
A coordinated observation was conducted among the ALMA, IRIS, and Hinode 
%The ALMA observation was coordinated with Hinode and IRIS from 13:36-19:41 UT 
on March 19, 2017. The ALMA 
%proposal numbered 
project, with project code 2016.1.00030.S (PI: Shimizu), was executed. 
The Hinode and IRIS operation plan was numbered as IHOP 327. 
We selected a tiny active region located in the eastern hemisphere on 
the solar disk as the observing region for the coordinated observations.
The region is seen as a compact bright region in the soft X-rays; it exhibits a magnetic
bipolar distribution on photospheric SDO/HMI magnetograms.

The ALMA observation was conducted at 100 GHz (Band 3) from 15:33 to 19:10 UT.
The observation consists of three runs (15:33--16:26 UT, 16:53--17:46 UT, and
18:16--19:10 UT); a soft X-ray microflare occurred during the first run.
This tiny active region is located at  ($-495$\arcsec, $-40$\arcsec) on the heliocentric coordinate
at the beginning of the first run. The antenna layout is the C40-1 configuration, which is the most compact
layout in the solar observing mode with sets of 39 and 7 antennas with diameters 12 m and 7 m, respectively.
%and 7 seven-meter
%antennas.  
The longest baseline is 249 m with a spatial resolution of $5.0  \times 3.9 \arcsec$.  
The common astronomy software applications %(CASA Ver 5.4: McMullin et al. 2007) 
\citep[CASA Ver 5.4; ][]{mcm07}
was used to calibrate and image
the data. The CLEAN procedure 'tclean' was used to synthesize each image 
\citep{shm17}.  The synthesis was conducted excluding the visibilities
between the two antenna sets (39 and 7 antennas) %the seven-meter antennas and twelve-meter antennas, 
because appropriate synthesis was not obtained when all the visibilities among the antennas
were used. 
When the $7-12$ m baselines were involved in the synthesis, artificial stripes 
appeared in the synthesized images. 
The stripes disappeared when the baselines were removed. 
We do not appropriately understand the origins of the stripes; however, 
the visibility data of $7-12$ m baselines degraded the image quality. 
Careful calibration might solve this issue; however, we removed the baselines in the image 
synthesis presented in this study because we do not know the calibration process for 
the $7-12$ m baselines. 
The data were self-calibrated in phase. % and amplitude. 
%To improve the image quality, we integrated the data acquired over 
%every 20 s. 
The self-calibration process has five steps with different accumulating periods to 
obtain the solution from 10 min for the first step to 2 s for the last step 
%We integrated the data acquired over every 2 s, 
such that the temporal cadence is 2 s in this study.
No single dish (total power) data were acquired to calibrate the absolute 
brightness temperature in this observation. Thus, the temporal profiles presented in
this article 
%we investigate only the temporal
%evolution of 
show
the deviation from the averaged brightness temperature in 
the observation field-of-view.  
The increase in temperature brightness  for the event considered
can be derived even without absolute temperature information.
%The primary focus of this article is time variability in ALMA data, namely, relative variability in
%temperature, which can be studied even without absolute temperature information. 

The IRIS satellite performed the spectroscopic and 
slit-jaw imaging observations during the period.
This article presents the results obtained from the slit-jaw images because the slit positions were
not present around the footpoint of the soft X-ray microflare.
The slit-jaw images were acquired at three channels 1400, 1330, and 2796, each of which
covers Si IV 1394/1402 \AA\ lines ($10^{4.8}$ K), 
C II 1335 \AA\ lines ($10^{4.2}$ K), and Mg II k line ($10^{4.0}$ K), respectively, 
at a cadence of 22 s. The pixel size is $0.34\arcsec$. 
Each image has a field of view of $60 \times 65\arcsec$, but it moves in the EW direction
at each exposure because of the scanning for spectroscopic measurements.
This provides the maximum coverage in the field of view, at $120 \times 65\arcsec$. The level 2
data from the IRIS were used in this analysis, which were obtained following 
instrumental calibration including dark current, flat field, and
geometric corrections.   %, available from the IRIS team, were used in this analysis.

The X-Ray Telescope (XRT) \citep{gol07, kan08} on the Hinode satellite \citep{kos07}
repeatedly and continuously acquired Al-poly filter images every 4 s with occasional
exposures of G-band images (XOB \#1B76). The image size is $128 \times 128$ pixels 
with a pixel scale of $1.031\arcsec$\citep{shi07}. The standard calibration routine on
Solar SoftWare was used to calibrate the time series of the XRT images.

The Hinode's Solar Optical Telescope (SOT) \citep{tsu08, sue08, shi08, ich08} 
repeatedly and continuously recorded 
the full-polarization states of line profiles of two magnetically sensitive Fe I lines 
at 630.15 and 630.25 nm using the spectro-polarimeter \citep[SP, ][]{lit13}.
The fast-mapping mode was used, which covers a $77 \times 80 \arcsec$ field of view
with an effective pixel size $0.32\arcsec$.
The spectral sampling is 21.549 m\AA\ pixel$^{-1}$. 
The map cadence was approximately 30 min.
We used the SOT/SP level 2 database%, which 
%consists of outputs from inversions using the MERLIN inversion code
%developed under 
%\edit1{ 
from%}
the Community Spectro-polarimetric Analysis Center (CSAC) 
%initiative 
%\edit1{
\footnote{http://www2.hao.ucar.edu/csac}%}
at HAO/NCAR%. 
%\edit1{
, which are results of% }
%The inversion code performs a least-squares
%fitting of the Stokes profiles using the 
Milne--Eddington 
%{\edit1{ 
inversions by the MERLIN code with
the calibration described in \cite{lit13b}.%}
%atmospheric 
%approximation, which allows for a linear variation of the source function 
%along the line-of-sight, but holds the magnetic field vector, line
%strength, Doppler shift, line broadening, and magnetic fill fraction as constant 
%along the line-of-sight. 

\subsection{Image Co-alignment and Spatial Distribution Correspondence}
%\subsection{Spatial Distribution Correspondence}
\label{sec:correspondence}
The data from the different instruments were carefully
investigated to identify a better method for the co-alignment; we finally achieved better 
than $2\arcsec$ in the uncertainty of the co-alignment. 
%, as shown below. %according to section \ref{sec:correspondence}. 
IRIS and Hinode data were mapped on the full Sun images from the 
Solar Dynamic Observatory \citep[SDO, ][]{pes12} by obtaining
the maximum correlation between the corresponding data; 
A Hinode SP map was mapped on the HMI
magnetogram \citep{sch12a, sch12b} by obtaining the best matching in the spatial distribution of 
the magnetic flux. The time series of the Hinode XRT images were mapped on an AIA 94 \AA\
coronal image to determine the co-alignment accuracy. 
The IRIS slit-jaw images were mapped on the AIA 1600 \AA\ 
chromospheric image using an image of 2794 \AA. 
For the ALMA data, 
the coordinates available for each map were used to co-align to the SDO full-sun images because the pointing accuracy of the ALMA antennas
has been established as more than $2 \arcsec$ \citep{alm16}.
%In addition to good correspondence in the pattern distribution to be discussed
%in  section \ref{sec:correspondence},
%we also had confidence for the good co-alignment (2") from the result that 
%a transient increase was observed in the ALMA exactly 
%at the footpoint location of a loop-like transient brightening observed with XRT 
%during this observational period. 
Nevertheless, the coordinates of the image provided by the ALMA observatory exhibit 
a problem when the coordinates are not appropriately revised using the reference time. 
Generally, the coordinates are not revised
%become less accurate 
when there is no pointing data 
that are derived from the encoders of the antennas. 
To avoid this problem, we revised the coordinate using the reference time, which 
differs from that mentioned in the readme file of the data package in the ALMA Archive.  

\begin{figure}[ht!]
\plotone{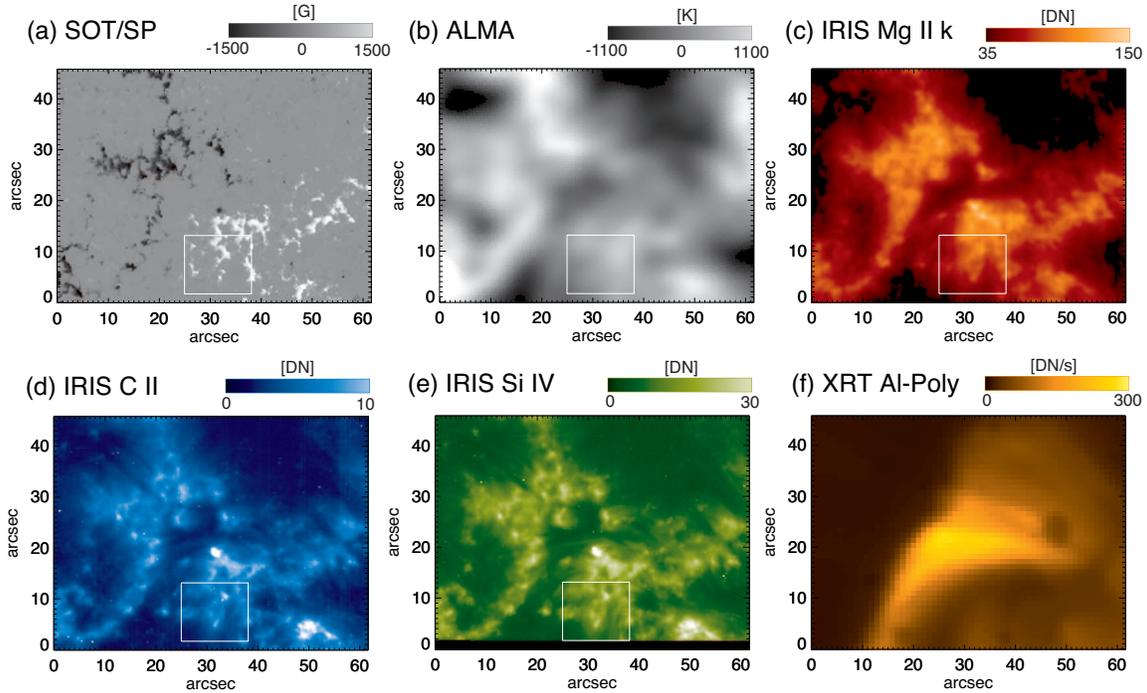}
\caption{Spatial distribution of time-averaged features visible in (a) SP line-of-sight magnetic flux density, 
(b) ALMA brightness temperature, (c) IRIS slit-jaw Mg II filter, (d) IRIS slit-jaw C II filter, 
(e) IRIS slit-jaw Si IV filter, and (f) Hinode XRT Al-poly filter. They are
co-aligned to each other according to the method described in section \ref{sec:obs}.
In (a), the white and black are positive and negative polarities, respectively. 
%The field of view
%shown here is the area examined in this article.
Solar north is upwards and west is to the right direction. 
%\edit1{
DN is data number units.%}
The boxes show the field of view of Figures~\ref{fig:siiv} and \ref{fig:diff}.
}
\label{fig:map}
\end{figure}

Figure~\ref{fig:map} shows the spatial distribution of features in the co-aligned images.
%seen in the field of view examined in this article. 
The SP line-of-sight magnetic flux density map
(Figure~\ref{fig:map}(a)) demonstrates a bipolar flux distribution in the field of view.
The areas where the magnetic flux is concentrated are brighter than their surroundings
in the three IRIS slit-jaw images (Figure~\ref{fig:map}(c)--(e)). 
Brighter features in the IRIS images correspond to small flux concentrations.
Diffuse features can be recognized around the bright features, which may be 
caused owing to the expansion
of magnetic flux at the chromospheric height. 
%Because the temperature responses of the Mg II, C II, and Si IV filters are 
%at peak values in the middle chromosphere ($10^{4.0}$ K),
%upper chromosphere ($10^{4.2}$ K) and a lower portion of the transition region ($10^{4.8}$ K), respectively,
%this comparison 
Bright features at flux concentrations
mean that high heating rate is sustained from the middle chromosphere to a lower portion
of the transition region at the magnetic flux concentrations;  
the temperature responses of the Mg II, C II, and Si IV filters are 
at peak values in the middle chromosphere ($10^{4.0}$ K),
upper chromosphere ($10^{4.2}$ K) and a lower portion of the transition region ($10^{4.8}$ K), respectively.
Figure~\ref{fig:map}(b) shows the corresponding ALMA brightness temperature map, which is time-averaged
over the first run.
% and co-aligned to the other maps according to the method
%described in section \ref{sec:obs}.  
The brightness temperature here is the deviation
from the temperature averaged over the field of view. 
The overall spatial distribution in the ALMA temperature map corresponds well
with those observed in the IRIS chromospheric maps.  
However, the ALMA map shows a more diffuse distribution, which may be because of
the lower spatial resolution of the ALMA in this observation.
Figure~\ref{fig:map}(f) is a co-aligned X-ray image.
The bright corona is not
observed as a bipolar structure connecting the flux concentrations to each other, which
is completely different from what is observed in the chromosphere and lower transition region. 

\section{Results} 
\label{sec:results}

\subsection{Loop-like Microflare in Soft X-rays}

\begin{figure}[ht!]
\epsscale{1}
\plotone{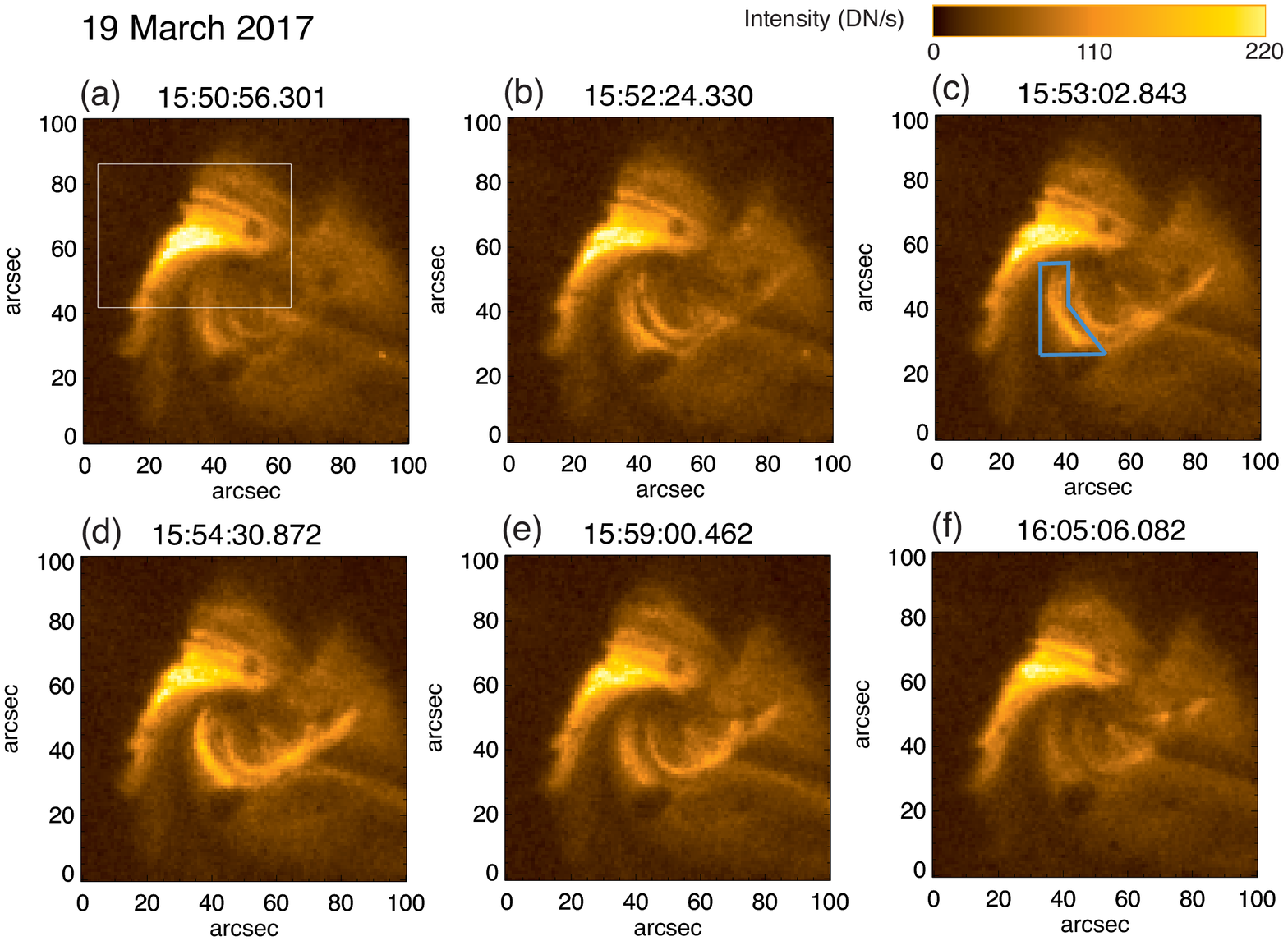}
\caption{
Time series of soft X-ray images (Al-poly filter) acquired by Hinode X-Ray Telescope.
%The order is from the upper left to the upper right and then from the lower left to the lower right.
The time sequence is from (a) to (c) and then from (d) to (f).
The field of view is wider than that in Figure~\ref{fig:map}. 
Solar north is upwards and west is to the right direction. 
The rectangle in (a)
%the 1st frame (upper left frame) 
shows the field of view used in Figure~\ref{fig:map}.
The area surrounded by light blue lines in (c)
%the 3rd frame (upper right frame) 
indicates the pixels
where the intensity was integrated for Figure~\ref{fig:lc}(a). 
}
\label{fig:xrt}
\end{figure}

\begin{figure}[ht!]
\epsscale{0.6}
\plotone{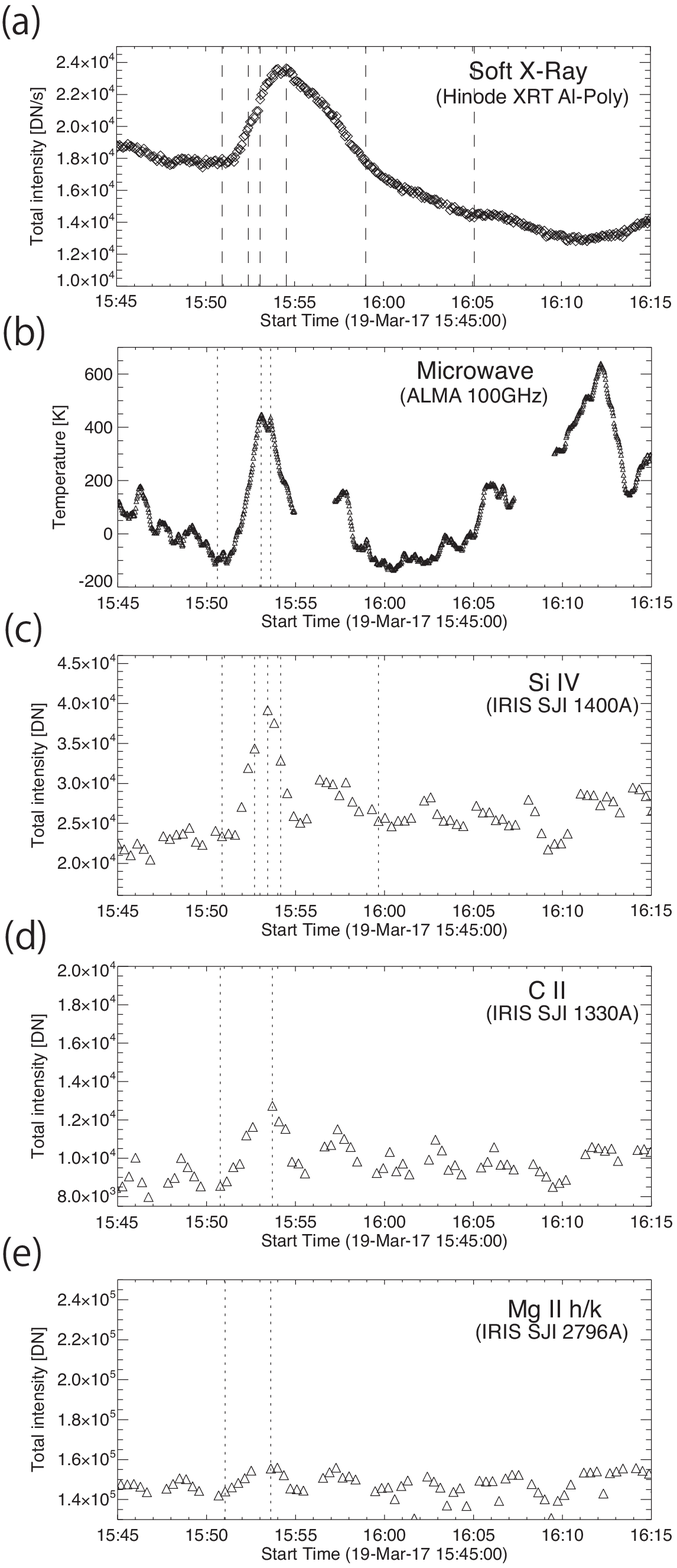}
\caption{Intensity profiles of (a) the soft X-ray microflare and its counterparts 
measured in (b) 100 GHz, (c) Si IV, (d) C II, and (e) Mg II k. 
The counterparts were measured at one of the footpoints of 
the soft X-ray brightening loop, while the soft X-ray profile is for the entire area
of the brightening loop, as specified in Figure~\ref{fig:xrt}. 
The vertical dash lines determine the time of the images shown in
Figures~\ref{fig:xrt}, \ref{fig:siiv} and \ref{fig:diff}. 
}
\label{fig:lc}
\end{figure}

Figure~\ref{fig:xrt} shows the time series of soft X-ray images
(with Al-poly filter) obtained from the Hinode XRT. 
Six frames were selected from the cadence series of images at 4 s
to show the occurrence of a loop-like microflare, which is located almost 
at the center of the images. 
A single coronal loop became apparent up to the 4th frame (15:54:41 UT),
which was the time when the intensity was at peak. 
This loop brightening started from 15:51 UT immediately after the 1st
frame, in which a couple of loop structures had been faintly visible previously.
Another loop structure existed at the right side parallel to the brightening loop considered;
it may have been extended towards the west direction. 
This structure showed a temporal evolution similar to that of
the brightening loop. 
Figure~\ref{fig:lc} (a) shows the time profile of the soft X-ray intensity for
the area identified with light blue lines in Figure~\ref{fig:xrt}.
The vertical dashed lines represent the time at which each frame 
in Figure~\ref{fig:xrt} was acquired. 
The intensity stayed at a constant level until 15:51 UT (1st frame in Figure~\ref{fig:xrt}),
followed by a transient increase in the intensity. 
The intensity required approximately 3 min
to attain its peak at 15:54:30 UT (4th frame in Figure~\ref{fig:xrt}).
The increased total intensity of the brightening loop is $5.8 \times 10^{3}$ DN s$^{-1}$,
%\edit1{
where DN is data number.%}
%which 
%\edit1{
It%}
was derived by integrating the pixels surrounded by the light blue lines in Figure~\ref{fig:xrt}(c). 
Then, the intensity gradually decreased and required approximately 10 min
to attain the minimum. This temporal evolution is conventional for the soft X-ray
behaviors observed well in the solar flares and microflares. 

One of the brightening loops, that is, the upper (north) end of the loop, is located within 
the observing fields of view of IRIS, ALMA, and Hinode/SP such that we can investigate
the behaviors of the footpoint at different wavelengths. 
Figure~\ref{fig:lc}(c) shows the area intensity profile in the Si IV slit-jaw channel of the IRIS.
The Si IV intensity began to increase at 15:52 UT and peaked at 15:53:25 UT,
followed by an immediate decrease of the intensity to the initial level
by 15:55 UT.  
%This transient intensity increase happened during the period when
%the soft X-ray intensity still increased and the Si IV increase ended before
%reaching the peak in soft X-ray. 
This transient intensity increase is faster than the ascending slope in the soft 
X-ray profile (Figure~\ref{fig:lc}(a)); the increased Si IV intensity quickly reduced before
reaching the peak in soft X-ray. 

\begin{figure}[h] %t!]
\epsscale{1}
\plotone{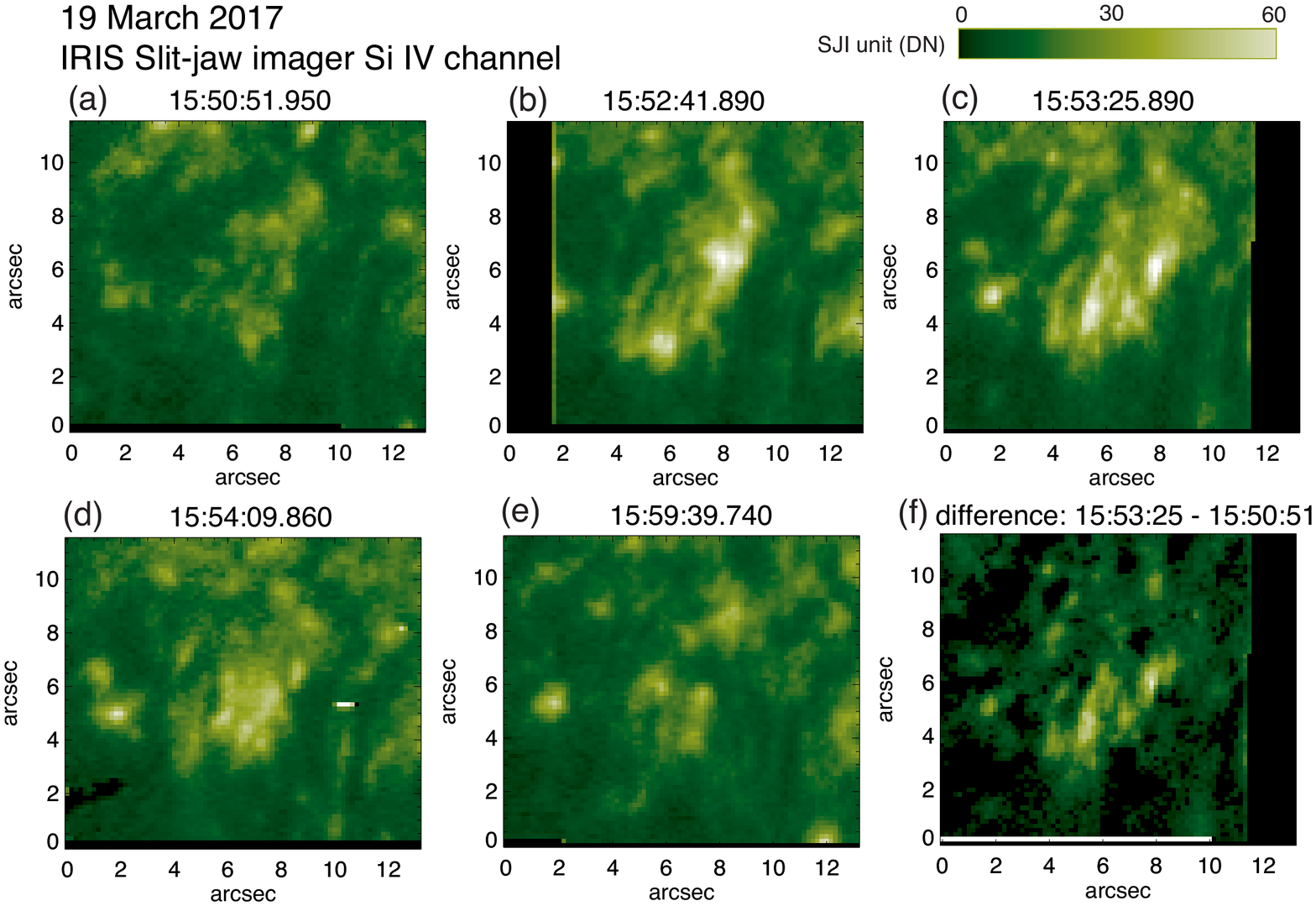}
\caption{Si IV channel images from the IRIS slit-jaw imager for the five times specified in
Figure~\ref{fig:lc}.  The 1st frame (a) is before the intensity began increasing,
and the 3rd frame (c) is at the intensity peak. The frame (f) is
obtained from the difference between (a) and (c). %  the 3rd and 1st frames. 
The field of view is $13.3 \times 11.6 \arcsec$
and its location is shown in Figure~\ref{fig:map}. 
}
\label{fig:siiv}
\end{figure}

Some Si IV images are shown in Figure~\ref{fig:siiv}, whose acquired times are
identified by the vertical dash lines in Figure~\ref{fig:lc}(c).
The field of view is $13.3 \times 11.6 \arcsec$, where the loop footpoint is located around the center. 
As described in Section~\ref{sec:obs}, a $60 \times 65 \arcsec$ field was repeatedly moved
in the EW direction at each exposure to sample spectral lines at several positions
in the $120 \times 65 \arcsec$ field of view. Because of this operation, observing field of view
changed with time, which can be partly recognized in the 2nd and 3rd frames as the dark area.
Four or five bright kernels, each of which has a size of less than $1 \arcsec$, 
appeared in the 2nd frame (15:52:41 UT). 
In the 3rd frame (15:53:25 UT), that is, at the peak time, 
the bright kernels showed a slight increase in intensity, location, and shape. 
However, we recognized two bright kernels at similar locations.
The frame at the lower right shows
the intensity difference in the 3rd frame when compared with the 1st frame (15:50:51 UT).
This clearly provides the location of brightening features at the footpoint of the soft X-ray microflare.
The bright features may be grouped into two areas centered 
at (X, Y) = (5.5, 4) and (8, 6) in arcsec.  
 
\begin{figure}[h] %t!]
\epsscale{1}
\plotone{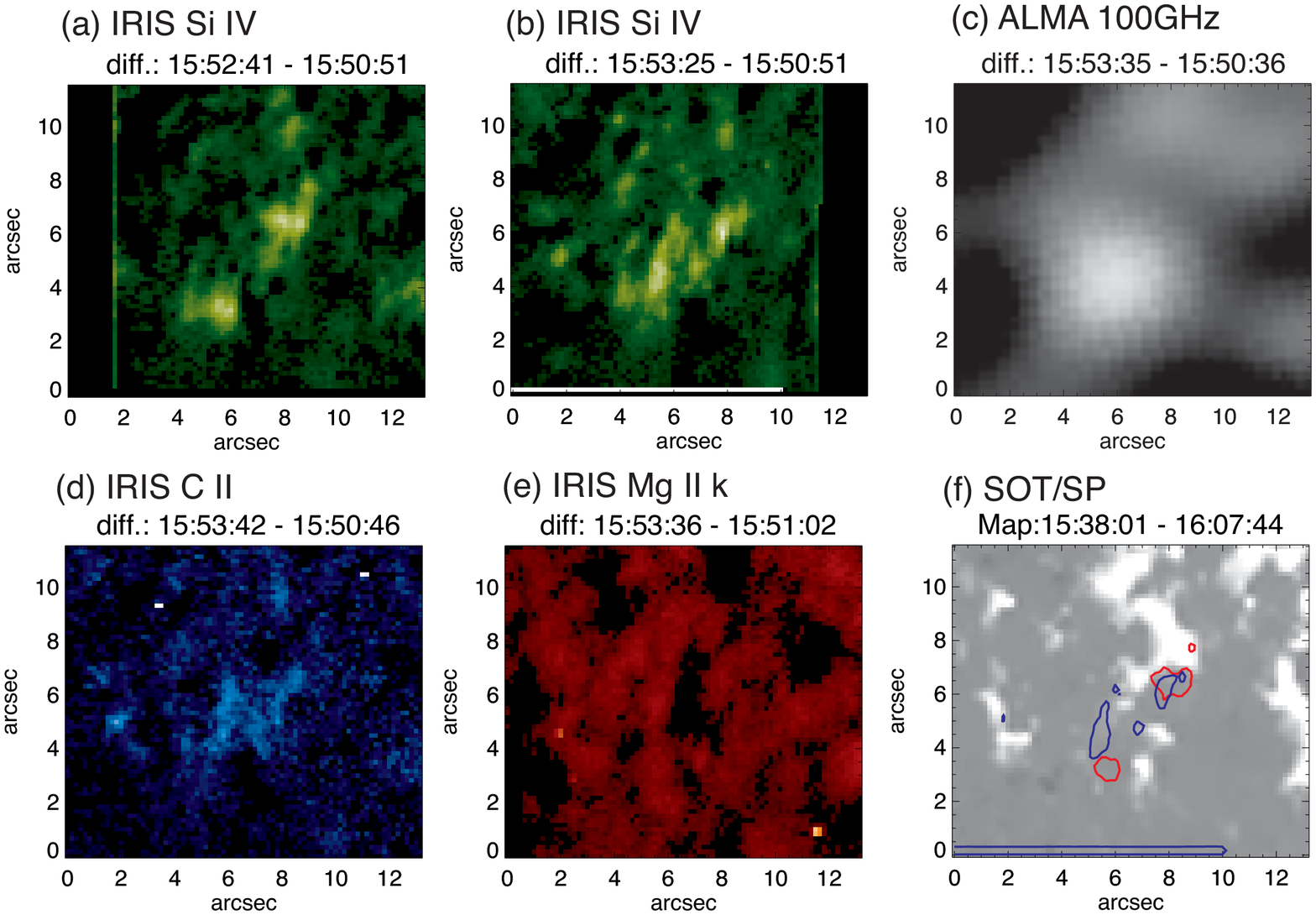}
\caption{
Difference frames for (a), (b) Si IV, (c) 100 GHz, (e) C II, and (f) Mg II k, which were 
obtained by subtracting the background image (identified by the 1st vertical dash line for each time
plot shown in Figure~\ref{fig:lc}). 
%(b) is identical to the frame at the lower right 
%in Figure~\ref{fig:siiv}. 
The frames in (b), (c), (d), and (e) are for the images recorded 
at peak time. (f) Spatial distribution of the magnetic flux density derived from the Hinode SP
scan in 15:38:01--16:07:44 UT. 
The red and blue contours indicate the position of Si IV bright kernels observed in (a) and (b), respectively.
The field of view is same as that of Figure~\ref{fig:siiv}.
}
\label{fig:diff}
\end{figure}

Figure~\ref{fig:lc}(b) is the temporal evolution of the brightness temperature
at 100 GHz for the region considered, which shows that the counterpart signals are detected
by ALMA at 100 GHz. 
The ALMA temporal profile of the brightness temperature is almost same as that in the Si IV channel. 
The intensity increase is also visible in the C II channel, as shown in Figure~\ref{fig:lc}(d),  
but not in the Mg II k channel (Figure~\ref{fig:lc}(e)). 
Figures~\ref{fig:diff} (b), (c), (d), and (e) show the Si IV, 100 GHz, C II, and Mg II k images, respectively, 
at the intensity peak time with the background removed.
The background is the frame specified by the 1st vertical dash line in each time
plot in Figure~\ref{fig:lc}. 
The counterparts are clearly detected in Si IV and 100 GHz at 15:52--15:53 UT.
The spatial resolution of the ALMA data is at least 10 times lesser than that of the IRIS. 
Although the intensity is weak, the counterparts can be observed in the C II frame. 
No apparent counterpart is observed in the Mg II k frame. 

Figure~\ref{fig:diff}(f) is the spatial distribution of magnetic flux at the photosphere. 
Almost all the bright kernels observed in Si IV and C II are located 
in weak, void areas surrounded by strong magnetic flux concentrations, as shown by the contours.
%No intensity increases are observed in the strong magnetic flux islands. 

\subsection{Thermal energy of the soft X-ray microflare}

\begin{figure}[h] %t!]
\epsscale{0.6}
\plotone{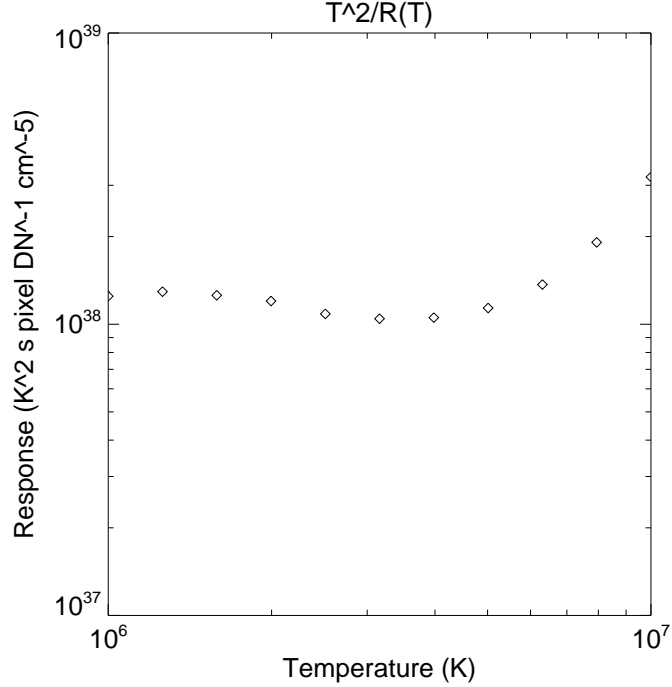}
\caption{
$T^2/R(T)$ as a function of temperature ($T$) for the XRT Al-Poly filter, where
$R(T)$ is the filter response curve.
}
\label{fig:response}
\end{figure}

Using the total soft X-ray intensity at the time of its peak intensity, 
we estimated the thermal energy content for the soft X-ray microflare. 
The thermal energy content $E_{th}$ is 
\begin{equation}
E_{th} = 3 n_{e} k_{B} T V, 
\end{equation}
where $n_{e}$ is the electron density, $k_{B}$ is the Boltzmann constant, 
$T$ is the electron temperature, and $V$ is the volume. 
In the Yohkoh era, many studies used a filter ratio method
for the temperature and density 
by assuming isothermal plasma 
\citep[e.g.][]{shi95, shm00},
%(e.g., Shimizu 1995, Shimojo and Shibata 2000), 
followed by more advanced techniques, such as 
differential emission measure.  In this observation, only Al-Poly filter images
were acquired to obtain dynamical behaviors with high temporal cadence.
Thus, we used the method proposed by \cite{sak14} to estimate the order of energy.
This method states that the thermal energy is not sensitive to the temperature
of the XRT filter intensities.
The filter response curve $R(T)$ is simplified to be
\begin{equation}
R(T)= \frac{I}{\int n_{e}^{2} dV} \approx \frac{I}{n_{e}^{2} V},
\end{equation}
where $I$ is the measured total soft X-ray intensity. Thus, the thermal energy content
can be described as
\begin{equation}
 E_{th} = 3 \sqrt {\frac{I}{R(T) V}} k_{B} T V
  = 3 k_{B} \sqrt {\frac{T^2}{R(T) }} \sqrt{I V}.
 %(\frac{I}{RV}) k_{B} T_{given} V
\end{equation}
The response curve of the Al-poly filter is approximately a power-law distribution 
as a function of temperature in $10^6 - 10^7$ K \citep{gol07}. 
If the power-law index is 2,  $R(T)$ is proportional to %\propto 
$T^2$. Thus, $E_{th}$ is insensitive to
the temperature. The Al-poly filter exhibits a power-law index approximately equal to 2, 
that is, it exhibits a very weak dependence on the temperature, as shown in
Figure~\ref{fig:response}. In our estimation, we used
$1.1 \times 10^{38}$ K$^2$ s pixel DN$^{-1}$ cm$^{-5}$ for $T^2/R(T)$, which is the average
of $4-8$ MK. Transient brightenings in active regions have conventionally exhibit
a temperature of $4-8$ MK \citep{shi95}. 
The $T^2/R(T)$ has an error range of $1.0 \times 10^{38} - 1.2 \times 10^{38}$
in the energy estimation. 
The volume of the brightening loop was determined by measuring the length ($l$ = 27") and 
width ($w$=3.7") of the loop. Assuming that the depth (the length of line-of-sight) of the loop 
is equal to the width, we simplified the volume $V$  as $l \times w^2$. 
The XRT pixel size (1") was used for error estimation.  This calculation obtained 
a volume of  $1.4 \times 10^{26}$ cm$^{3}$
with an uncertainty range between $0.75 \times 10^{26}$ cm$^{3}$  and $2.4 \times 10^{26}$ cm$^{3}$.
Therefore, the thermal energy content was estimated to be
$2.9 \times 10^{26}$ erg with an uncertainty range between
$2.0 \times 10^{26}$ erg and $4.0 \times 10^{26}$ erg. 
The thermal energy produced by the microflare in the corona is larger than the thermal 
energy content because the thermal energy content is the amount of the thermal energy 
at a single time (peak time) without
considering energy losses from the radiation and thermal conduction during the ascending phase
\citep{shi95}.

\subsection{Thermal energy measured at the chromosphere}
\label{sec: chromoE}

The ALMA data can provide an order estimate of the thermal energy supplied to the chromosphere
to which the 100 GHz microwave is sensitive.  
Here, we assumed that the ALMA signal is the thermal free-free radiation. 
As observed from Figure~\ref{fig:lc}, the ALMA signal was peaked during 
the ascending phase of the soft X-ray emission. 
No polarization was measured using ALMA in this observation; thus, we cannot
ignore gyro-cynchrotron radiation for the transient enhanced signal at the footpoint
in the ALMA data.
Nevertheless,  we interpreted it as the signal caused by the thermal plasma
heated in the chromosphere by the electrons that are predominantly impinging 
the chromosphere.

The thermal energy supplied to the chromosphere is evaluated as 
the thermal energy content $E_{ch}$, which is 
\begin{equation}
E_{ch} = 3 n_{e} k_{B} \Delta T V, 
\end{equation}
where %$n_{e}$ is electron density, $k_{B}$ Boltzmann constant, 
$\Delta T$ is the excess in electron temperature. %, and $V$ volume
\citep{nin20}. 
Because the brightness temperature is a reasonable measure of the gas
temperature at an effective formation height in the chromosphere \citep{lou15}, 
for the increase in electron temperature, we obtained the increase 
in brightness temperature above the background level at the time
when the ALMA time profile was at its peak.  
The ALMA application (CASA) derived the flux density defined as 
the integration of brightness over the antenna beam angle 
in the units of Jansky per beam size. 
Assuming that the beam is Gaussian, the flux density was converted to 
brightness temperature, according to the formula presented in ALMA Cycle 4 
Technical Handbook. 
The background level is the brightness temperature just before
the event, whose time is identified by the left most dashed line  
in Figure~\ref{fig:lc}(b). The increased brightness temperature derived
was 545 K. 

The electron density was assumed to be in the order of $10^{11}$ cm$^{-3}$ 
according to certain numerical modeling efforts. \cite{lou15} derived
that the range of brightness temperature for an active-region magnetic structure 
is from 2,945 K to 13,104 K with an average of 6,090 K and root mean square 
variation of 1,333 K for the 100 GHz radiation. The 100 GHz (3 mm) radiation has
effective formation height of approximately 1,500 km with a contributing range of  
$1,200 - 2,000$ km depending on the magnetic structures, where the electron
density is approximately $10^{11}$ cm$^{-3}$.

The volume used in the evaluation was determined by assuming that the area where 
the 100 GHz radio emission is in excess during the event was almost same as the footpoint area of 
brightenings clearly detected in the Si IV (Figure~\ref{fig:siiv}).  
The upper left frames in Figure~\ref{fig:diff} indicate that the footpoint area changed
from image to image. For simplicity, we selected $8\arcsec$ for the N-S direction 
and $5\arcsec$ for the E-W direction
to cover the entire area of various footpoints $A$ with $\pm 1 \arcsec$ for uncertainty evaluation. 
Assuming that the extent in the height is comparable to the horizontal size, the volume $V$
was estimated to be $A^{3/2}$. This might be an overestimation because the effective
height of the 100 GHz radio is 800 km (approximately $1\arcsec$).
However, the transient heating by impinging electrons
at the chromosphere may expand the heated plasma towards the upper layers. 
With these parameters, we obtained $(2.2 \pm 1.2) \times 10^{24}$ erg as
the thermal energy supplied to the chromosphere.  

\section{Discussions} 
\label{sec:discuss}

In this article, we investigated the heating of the footpoints using 
IRIS and ALMA when an energy of $2.9 (\pm 1.0) \times 10^{26}$ 
erg was transiently released  in coronal structure that is observed 
as a single brightening loop using the Hinode XRT. 
Counterparts were clearly detected in the Si IV and ALMA at 100 GHz.
 They are visible but less apparent in the C II, and it is difficult to 
 distinguish them in the Mg II k. 
Their time profiles were impulsive; the intensities attained their peaks at approximately 1.5 min
before the peak of the soft X-ray intensity was attained. 
This was followed by a prompt decrease
in intensity in a few minutes. 
This temporal behavior is similar to the Neupert effect \citep{neu68}, which
has been observed in (non-thermal) hard X-rays 
pertaining to the soft X-ray (thermal) temporal behavior.
Moreover, the exact position of the footpoints was identified with 
high spatial resolution maps of the magnetic flux density at the photosphere
from the Hinode SOT/SP with a spatial relation to bright kernels in Si IV. Almost all
the bright kernels are located in weak, void areas surrounded by strong magnetic
flux concentrations. 

\subsection{Height dependence}
\label{sec: height}

The Si IV (1400 \AA\ ) filter is dominated by the Si IV 1394/1402 \AA\ lines with the continuum.
The continuum is sensitive to the upper photosphere and low chromosphere, but
the Si IV lines are formed at an equilibrium temperature of approximately 80,000 K, 
which corresponds to
the temperature in the transition region \citep{pet14}.
The C II (1330 \AA\ ) filter is dominated by C II lines at 133.5 nm (upper chromosphere) 
with the continuum while the Mg II k (2796 \AA\ ) filter dominates the Mg II k (chromosphere) 
and inner wings (photosphere). The C II and Mg II k lines are formed at similar heights
below the transition region, but the C II lines can be formed at a height higher and lower than 
the core of the Mg II k line depending on the quantitative measure of the materials 
in the temperature range of 14,000 -- 50,000 K.
Significant amounts of materials in this temperature range
form the C II lines at a height higher than that of the Mag II k line core \citep{rat15, rat15b}. 

Certain studies derived the mean brightness temperatures (with standard deviation) 
of the 3 mm ALMA band 7,300 ($\pm 100$) K \citep{whi17}, 
7,250 ($\pm 170$) K \citep{ali17} for the disk center, and 7,250 ($\pm 1,140$) K \citep{das18} 
for radiative magnetohydrodynamics (MHD) simulation of an enhanced network region.  
 \cite{das18} studied the diagnostic potential of combined optical, ultraviolet, and ALMA observations
and showed that the response function to temperature is at maximum at log $\tau = -5.6$, distributed
between log $\tau = [-6.0, -5.1]$,  which is slightly higher or almost same as that of the Mg II k line core, 
that is, the maximum at log $\tau = -5.5$, distributed between log $\tau = [-6.4, -5.1]$, 
where $\tau$ is optical depth%.
%\edit1{ 
at 500 nm.%} 

From these studies, we can interpret that the order of the mean temperature is 
Si IV, C II, 100 GHz, and Mg II k from higher to lower temperatures. Furthermore,
C II, 100 GHz, and Mg II k are distributed in a small temperature range at the upper chromosphere
but below the transition region. Considering the temperature distribution models of the solar atmosphere
as a function of height, the Mg II k image is located at the lowest height, followed by
100 GHz, C II, and Si IV toward the upper heights.   
This order of heights, with particles impinging from the corona to the chromosphere,  
can explain the observation at one of the footpoints
for the soft X-ray microflare, that is, more apparent signals were observed
at Si IV, 100 GHz, C II, and Mg II k, respectively.

%\edit1{
\subsection{Thermal energy supplied to the chromosphere}
\label{sec: reliable}
%}

%\edit1{
In Section~\ref{sec: chromoE}, the thermal energy supplied to the chromosphere was 
estimated on the assumption of thermal free-free radiation.
The increased brightness temperature from the ALMA signal 
was used for the excess in electron temperature.
This is valid when the optical thickness is unity or larger, i.e., optically thick,
at 100 GHz. 
The relationship between brightness temperature and electron temperature can be
written as
\begin{equation}
T_{b} = T_{e} (1 - e^{-\tau_{100}}),
\label{eq: temp}
\end{equation}
where $\tau_{100}$ is the optical thickness at 100 GHz \citep{kra86}. 
%\approx \tau_{m} T_{e},
As the optical thickness decreases, the brightness temperature becomes
smaller than the electron temperature. For the optical thickness 
much smaller than the unity, the brightness temperature can be approxiated by
$\tau_{100} T_{e}$.
%Thus it's important to check the optical thickness of the atmospheric layer
%observed with ALMA.
%}

%\edit1{
\cite{ali20} measured the brightness temperature from ALMA
center-to-limb observations at 100 GHz for magnetic bright networks.
% and cell interiors. 
From the inversion of this measurement, 
the electron temperature was derived as a function of 
the optical thickness at 100 GHz 
for magnetic bright networks, giving
$7,383-7,916$ K  
where the optical thickness is unity at 100 GHz.
They also computed the electron temperature as a function of
the optical thickness using opacity from the model parameters
of Model F, characterized as very bright network element, of 
the FAL93 \citep{fon93} model. 
The Model F gives the electron temperature of 9,200 K 
for the optical thickness of unity at 100 GHz. 
 These values are only $0-25$ \% increase from the mean brightness temperatures 
 at 100 GHz (Section~\ref{sec: height}) and thus equation~(\ref{eq: temp}) provides
that the optical thickness is about unity or higher. With this assessment, 
we conclude no big difference for the order estimation of thermal energy supplied 
to the chromosphere event if the brightness temperature is used as 
the electron temperature.
Similar treatment is applied by \cite{nin20, nin21}, which derived the energy 
of transient brightenings observed in the quiet Sun with ALMA.
%}

%\edit1{
We assumed thermal free-free radiation for the signals detected with ALMA. 
\cite{kon18} examined mm and UV radiations from bright flare ribbons observed
at various M- and X-class flares and showed that mm radiation (including 100 GHz) 
is consistent with free-free radiation from the plasma of flare ribbons at
temperatures $10^4-10^6$ K, providing the assumption is reasonable
for the order estimation of a much smaller energy releasing event examined
in this study. However, the plasma may be in a dynamic state due to the transient 
response of the flux of impinging particles, which can deviate largely 
the mm radiation from what is expected by thermal free-free radiation.
As \cite{mor20} points out, it is important to consider not only thermal flare plasma 
emission but also dynamics of accelerated electrons and nonequillibrium effects on 
radiation transfer. 
Better understanding of these effects would predict the intensity of the radiation
at 100 GHz more accurately and improve the accuracy of the energy estimation.
Such efforts will enhance the validity of the energy obtained in this study.
%}

\subsection{Energy partition}
\label{sec:partition}
The thermal energy supplied to the chromosphere was estimated at
$(2.2 \pm 1.2) \times 10^{24}$ erg with the 100 GHz measurement.
This is
only 1\%  of the thermal energy observed at the peak time of the soft X-ray
brightening loop in the corona, $(2.9 \pm 1.0) \times 10^{26}$ erg. 
It is only 2\% even when the same amount of energy was transported to the two
footpoints of the coronal loop. Considering the temporal evolution of 100 GHz
signals with respect to the soft X-ray thermal evolution, the thermal energy from
the 100 GHz measurement is created by the particles impinging the dense
layers. Thus, this estimated energy can be regarded as the energy exhibited by 
non-thermal accelerated particles required for the order estimation. 
The energy partition between the thermal and non-thermal plasmas in solar flares 
has been an important issue in the physics of magnetic reconnection
and particle acceleration. For example, \cite{ems04} evaluated the energy budget for
two X-class flares/CME events and showed that 
the peak energy in the thermal soft X-ray plasma (in the order of $10^{31}$ erg) 
is approximately one order of magnitude 
less than the energy in the accelerated particles. 
\cite{sai02} also evaluated that
the thermal energy of the flare kernel (approximately $10^{29}$ erg) 
was more than an order of magnitude smaller than
the electron beam energy for a smaller (C9.7 class) compact flare.
More recent studies reported the energy partition for more numbers of flares
\citep{sto07, ems12, ing14, war16a, war16b, ash17}. However, they provided
contradicting results on energy partition, particularly for weak flares. 
For A- and B-class flares, 
%corresponding
%to the larger energy group of microflares, 
\cite{ing14} derived the ratio of nonthermal energy and
peak thermal energy at a few times of $10^{-1}$ for 10 events, 
whereas \cite{sto07} obtained $10^{1} - 10^{2}$ from 18 events.  
\cite{war20} reviewed these studies and identified uncertainties in deriving the energy partition.   
Considering these studies together with the identified uncertainties, they suggested
that the thermal--nonthermal energy partition changes with the flare importance%.
%In weak flares, there appears to be 
%\edit1{
; a deficit of energetic electrons is observed in weak flares,%}
while the %injected 
nonthermal energy is sufficient to account for the thermal component in 
strong flares. 
However, we did not evaluate the energy partition for much smaller events.

Our observation detected one microflare with various instruments 
to discuss the energy partition. However, it is only one event.  
Our result may suggest that energetic electrons are much less generated when the magnitude
of energy release events decreases to the small energy group of microflares, that is, 
$10^{26}$ erg. It is in the order of $10^{-2}$ in the case of the event examined. 
Less number of energetic electrons may also mean that the heights that the energetic electrons
can impinge to are not low. \cite{bat11} showed that the hard X-ray emission at 30 keV  is 
approximately 1,000 km above the photosphere. Moreover, the energy deposition rate as a function of height 
for two different cases of electron cutoff energy shows that the electrons of $12-20$ keV
predominantly release the energy at $2,000 - 3,000$ km.
These heights correspond to the upper chromosphere or
lower transition region with a density of 
between $10^{11}$ cm$^{-3}$ and $10^{12}$ cm$^{-3}$, where the Si IV line
can be formed when compared to radio at 100 GHz, C II and Mg II k lines.   

Energy deposition at the chromosphere by a beam of energetic electrons 
drives strong chromospheric evaporation leading to a significantly denser 
corona and much brighter emission in the corona. 
Flare model simulations indicate that the energy flux 
of an electron beam required for explosive evaporation is 
approximately $2 \times 10^{9}$ erg cm$^{-2}$ s$^{-1}$ for 10 keV electrons 
and increases to $10^{10}$ erg cm$^{-2}$ s$^{-1}$
for 20 keV electrons \citep{ree15}. 
Because two bright kernels with a size of approximately $1 \arcsec$ were observed in Si IV,
the area to which electron beams impoinge is approximately $1 \times 10^{16}$ cm$^{2}$.
Assuming that 10 s is the duration at which electron acceleration occurs, 
the beam energy threshold for turning on explosive evaporation is estimated 
to be $2 \times 10^{26}$ erg, which is two orders of magnitude
smaller than that of the energy deposition measured at the chromosphere. 
Thus, explosive evaporation did not work in this event, resulting in
less amount of the heated plasma supplied to the coronal loop structures. 
Considering the temperature response curve of the XRT Al-poly filter, 
the magnitude of the observed soft X-ray increase can be expected
with a $1-2$ MK increase of temperature even when less amount of 
heated plasma is supplied from the chromosphere. 
The soft X-ray morphological evaluation (Figure~\ref{fig:xrt}) showed
no transient brightening at the loop footpoints. 
%Since the thermal energy is dominantly generated by the energy release in the corona, 
%the coronal plasma in the magnetic loop structures is more efficiently heated. 
%If the energy release site would be located at the apex of the loop structures, we would 
%observe the plasma heating, i.e., intensity increase, at the apex. 
%A deficit of accelerated particles impinging to the footpoints does not cause intense
%transient heating at the footpoints, i.e., chromospheric layers, and poor
%evaporation of chromospheric plasma into the coronal loop structures.
%As a result, we would not observe transient brightening starting from the loop footpoints.
%More than a half of soft X-ray microflares showed brightening from one or two footpoints
In soft X-ray microflares, the brightening loops 
brighten from their footpoints, followed by the brightening of the entire loop \citep{shi94}.
However, in this event, the brightening loop gradually brightened
from the apex without brightening at the footpoints, followed by
the brightening of the entire loop. 

\subsection{Magnetic environment at the footpoint and configuration for energy release}

The exact location of the soft X-ray brightening loop was identified by transient
brightenings in high resolution Si IV images. 
The Si IV brightenings consist of more than four bright kernels surrounded by
diffuse intensity increase. 
The bright kernels represent the location of magnetic flux 
in which accelerated particles generated in the corona impinged 
the transition region. 
%Figure~\ref{fig:diff}(f) clearly shows that the bright kernels are located in weak
%and void areas in magnetic flux at the photosphere. 
%This indicates that the magnetic effects like mirror effect are not 
%the reason why a deficit of accelerated particles is observed at the footpoint
%of this event.
Although the distance of the change is less than $1 \arcsec$, 
the position of the bright kernels changes with time frame by frame. 
This suggests that particle acceleration occurs in the magnetic field
lines connecting several positions at the low atmosphere, and 
that the magnetic field lines associated with particle acceleration
change with time. 
Several positions of magnetic field discontinuity
exist in the magnetic structure relating to the microflare in the corona 
if the energy release shows magnetic reconnection.

Comparison of the Si IV brightening kernels 
to the photospheric magnetic flux map (Figure~\ref{fig:diff}(f)) showed that
all the brightening kernels are located in weak magnetic areas
surrounded by strong magnetic flux distribution.
This means that the heated coronal magnetic structures are rooted
into the weak magnetic fields at the photosphere. 
In such magnetic environment, the magnetic effects such as mirror effect do not work
and thus, they did not cause 
a deficit of accelerated particles observed at the footpoint
of this event.

This field connectivity is reasonable considering pressure balance; 
as the gas pressure is higher in the brightening magnetic bundle than
outside, the pressure balance exhibits lower magnetic pressure in
the bundle. Thus, the magnetic flux density of the brightening loop
structure may be lower than that of the surroundings.  
When the brightening loop structure is formed with magnetic reconnection,
the field discontinuities for the reconnection may be located between the magnetic
field lines arising from the strong magnetic islands just beside
the Si IV brightening kernels. 
%\edit1{
As the pressure of external longitudinal magnetic fields changes, 
unstable current sheets may be formed preferably 
around and along twisted magnetic flux ropes, as investigated by 
\cite{tsa20} and \cite{sol21}.  
%}

\section{Summary} 
\label{sec:summary}

We investigated the footpoint behaviors of a soft X-ray loop-type microflare, which
was captured during an ALMA campaign coordinated with
Hinode and IRIS instruments. 
Counterparts at one footpoint of a brightening coronal loop were detected
in Si IV slit-jaw images and ALMA at 100 GHz, while they were less apparent in the C II and Mg II k
images. Their counterparts showed impulsive temporal profiles, which were at peak during
the rising phase of the soft X-ray intensity, followed by the intensity backing to the pre-brightening level
at the peak time of the soft X-ray intensity. 
The magnitude of thermal energy measured with ALMA is approximately 100 times smaller 
than that measured in the corona. 
These results suggest that impulsive counterparts can be detected in transition region 
and upper chromosphere at the footooints when soft X-ray loop-type microflares 
occur in the corona. 
The impulsive counterparts are considered as signals of thermal plasma heated at 
the upper chromosphere or transition region by impinging non-thermal particles,
although we should consider the addition of polarization measurements in future.
The energy evaluation indicates a deficit of accelerated particles that impinge the footpoints
of a small class of soft X-ray microflares. 
The footpoint counterparts consist of several tiny brightening kernels, all of which are 
located in weak and void magnetic areas formed in strong magnetic flux patchy distribution 
at the photospheric level. They provide a conceptual image in which the transient energy 
release occurs at multiple numbers of discontinuity formed on the sheaths of magnetic flux bundles
in the corona. 

To the best of our knowledge, the event presented in this article is 
a unique case in which the ALMA, IRIS, and Hinode instruments 
captured the foopoint behaviors of a small class of soft X-ray loop-like microflares.
To understand if footpoint behaivors presented in this article are common in
microflares, further observations should be conducted. 
%We are hoping more cases in coming years.  

\bigskip

This study used the following ALMA data: ADS/JAO.ALMA\#2016.1.00030.S. 
ALMA is a partnership of ESO (representing its member states), NSF (USA) and NINS (Japan), 
together with NRC (Canada), MOST and ASIAA (Taiwan), and KASI (Republic of Korea), 
in cooperation with the Republic of Chile. 
The Joint ALMA Observatory is operated by ESO, AUI/NRAO and NAOJ.
Hinode is a Japanese mission developed and launched by ISAS/JAXA, with NAOJ
as domestic partner and NASA and STFC (UK) as international partners. It is
operated by these agencies in cooperation with ESA and NSC (Norway). 
The Hinode/SP level 2 data (10.5065/D6JH3J8D) are provided by
the Community Spectropolarimetric Analysis Center at HAO/NCAR.
IRIS is a NASA small explorer mission developed and operated by LMSAL
 with mission operations executed at NASA Ames Research center and major 
 contributions to downlink communications funded by ESA and the Norwegian Space Center. 
This work was partially supported by JSPS KAKENHI Grant Number 15H05750, 15H05814, 
and 18H05234.

\end{document}